\documentclass[reprint,onecolumn,superscriptaddress,amsmath,amssymb,showpacs,prl]{revtex4-1}
\usepackage{hyperref}
\pdfoutput=1 

\usepackage{dcolumn}
\usepackage{epsfig}
\usepackage{graphics}
\usepackage[latin1]{inputenc} 
\usepackage[T1]{fontenc}
\usepackage{bm}
\usepackage{hyperref}
\newcommand{\ddst}{false}

\begin{document}

\title{Topological Constraints and Rigidity of Network Glasses from Molecular Dynamics Simulations}

 \author{Mathieu Bauchy}
 \email[Contact: ]{bauchy@ucla.edu}
 \homepage[\\Homepage: ]{http://mathieu.bauchy.com}
 \affiliation{Physics of AmoRphous and Inorganic Solids Laboratory (PARISlab), Department of Civil and Environmental Engineering, University of California, Los Angeles, CA 90095, United States}
 
\date{\today}


\begin{abstract}
Due to its non-crystalline nature, the glassy state has remained one the most exciting scientific challenges. To study such materials, Molecular Dynamics (MD) simulations have been extensively used because they provide a direct view into its microscopic structure. MD is therefore used not only to reproduce real system properties but also benefits from detailed atomic scale analysis. Unfortunately, MD shows inherent limitations because of the limited computational power. For instance, only the simulations of small systems are currently permitted, which prevents from studying small compositional changes, although it is well known that they can dramatically alter system properties. At this stage, it is tempting to follow topological constraint theory, which aims at describing macroscopic properties of the glass relying only on the connectivity of individual atoms, thus considering the complicated glass network as simple mechanical trusses. Thanks to only basic hand calculations, this theory has been successful in predicting complex composition and temperature behavior of glass properties, such as the glass forming ability, the viscosity of the corresponding liquid or the elasticity. The purpose of my PhD work is to connect these successful approaches based only on the topology of the underlying low-temperature network with properties that can be obtained from MD calcultations. It should thus allow for an increased applicatibility of rigidity theory.
\end{abstract}

\maketitle

\section{Topological constraint theory of glass}

According to this theory, pionnered by Phillips and Thorpe in the 80's \cite{phillips_topology_1979, thorpe_continuous_1983}, the rigidity of a glass can be evaluated by counting the number of constraints experienced by the individual atoms. In a covalent glass network, one needs to take into account both two-body radial bond-stretching (BS) and three-body angular bond-bending (BB) constraints. Phillips predicted that the glass-forming tendency is maximized when the number of mechanical constraints per atom ($n_c$) equals the number of degrees of freedom per atom ($n_d=3$), a condition which corresponds exactly to the Maxwell stability criterion for trusses. The corresponding network is then termed as \textit{isostatic}. If the number of constraints per atom is less than 3, the network is considered as being \textit{flexible} and shows low frequency (floppy) deformation modes. On the other hand, if the number of constraints per atom is greater than 3, the network is considered \textit{over-constrained}. \cite{mauro_topological_2011}

In covalent systems, such as in chalcogenides, the mean coordination number $\overline{r}$ is the key control parameter and evaluating the number of BS and BB constraints per atom is quite straightforward (one has $\overline{r}/2$ BS and $2\overline{r}-3$ BB). The total number of constraints per atom is thus given by :
\begin{equation}
 n_c = \frac{\overline{r}}{2} + 2\overline{r} - 3
 \label{eq:threshold}
\end{equation} so that the Maxwell stability criterion ($n_c = 3$, the rigidity percolation threshold), occurs at $\overline{r} = 2.4$. This theory has successfully predicted the rigidity percolation threshold of numerous chalcogenide systems and has now been also extended to oxide glasses.

Although the original Phillips-Thorpe theory predicts a single threshold where the network fulfills $n_c = 3$, recent modulated differential scanning calorimetry experiments by Boolchand and co-workers \cite{feng_direct_1997} have revealed the existence of a reversibility window defining more than one simgle isostatic composition. This has led to the recognition of an intermediate phase (IP), defined between the usual flexible and stressed-rigid phases. The existence of the IP has been attributed to a self-organization of the network with increasing stress (connectivity) as illustrated through a phenomenological cluster model. \cite{micoulaut_onset_2007}

\section{Relating Molecular Dynamics to constraints}

\subsection{General method}

Even if the applicability of the topological constraint theory is rather simple when atoms follow the 8-N (octet) rule, one has to be careful on how to use it. Especially, the link between coordination number (CN) and constraints and the distinction between soft and hard constraints need to be carefully considered. For example, for stoichiometric compositions like GeSe$_2$ or GeS$_2$, a simple counting leads to $n_c = 3.67$ (i.e. an over-constrained glass) in agreement with the fact that these compositions exhibit a low glass-forming ability. Applying the same algorithm to the corresponding oxides (GeO$_2$ and SiO$_2$) would lead to the same value of $n_c$, which would actually be challenged by experiment indicating that silica and germania form glasses rather easily. In silica, the constraint associated to the inter-tetrahedral angle Si-O-Si is actually broken so that its network is isostatic. \cite{zhang_central_1994} Moreover, if constraint counting in tetrahedral glasses is rather straightforward, alkali silicate systems are more challenging since alkali atoms do not follow the 8-N rule. For instance, the CN of sodium atoms is found to be 5 from MD simulations and EXAFS experiments \cite{houdewalter_sodium_1993}, although it has been proposed that only one BS constraint is relevant. These two examples illustrate that an exact knowledge of which constraint should be taken into account is needed in order to apply rigidity theory with confidence, and it can certainly not be obtained by simple guesses on CN or interactions.\medskip

We have therefore proposed to use Molecular Dynamics in order to compute the number of constraints per atom at the atomic scale, without any prerequisite on coordination number or interactions. \cite{bauchy_angular_2011, bauchy_atomic_2011} In valence-force-field models, an atom experiencing bond-stretching ($\alpha$-type) and bond bending ($\beta$-type) interactions is characterized by the total potential energy $U$ given by $U = (1/2)k_{\alpha}\sigma_r ^2 + (1/2)k_{\beta}r^2\sigma_{\theta}^2$ where $\sigma_r$ and $\sigma_{\theta}$ are the standard variations of respectively the bond length $r$ and the bond angle $\theta$. Thus, as in classical mechanics, one can treat potential energies or forces and deduce the motion. However, one can perform the opposite and relate atomic motion to the absence of restoring forces which would maintain bonds and angles fixed around their average value. It is thus equivalent to evaluate the rigidity of a constraint by calculating its potential energy or by computing the standard deviation ($\sigma_r$ for BS and $\sigma_{\theta}$ for BB) of the corresponding atomic motion, which give rise to radial or angular distributions.

\begin{figure*}
\begin{center}
\includegraphics*[width=0.5\linewidth, keepaspectratio=true, draft=\ddst]{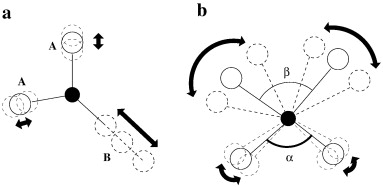}
\end{center}
\caption{Method of constraint counting from MD-generated configurations. Large (respectively small) radial (a) or angular (b) excursions around a mean value are characterized by large standard deviations on bond B (respectively small on A) representing broken (respectively intact) constraints.}
\label{fig:motion}
\end{figure*}

\subsection{Bond-Stretching (BS) constraints}

\begin{figure*}
\begin{center}
\includegraphics*[width=0.6\linewidth, keepaspectratio=true, draft=\ddst]{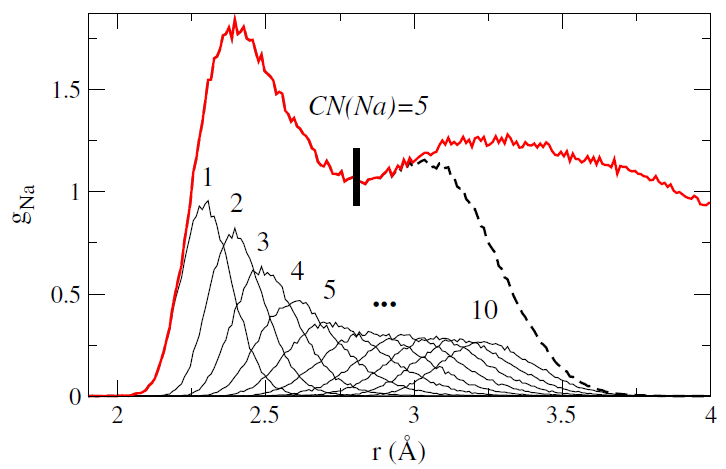}
\end{center}
\caption{Na pair distribution function $g_{Na}$ in the NS2 glass and its decomposition into 10 neighbor distributions out of which are computed correspoding radial standard deviations $\sigma_r$. The broken curve is the sum of the 10 distributions.}
\label{fig:BS_method}
\end{figure*}

During the course of the PhD work, we have first focused on three different alkali disilicate systems 2SiO$_2$-M$_2$O with M = Na, K and Li (referred as NS2, KS2 and LS2). Having generated atomic scale configurations for each system, pair distribution functions have been computed and splitted into neighbors contributions (1, 2,...6). The standard deviations $\sigma_r$ of each neighbor distribution give an estimate of the strength of the corresponding BS constraint, i.e. a large $\sigma_r$ will be associated to a broken constraint indicative of large radial excursions whereas a small standard deviation will be associated with an intact constraint (see Fig. \ref{fig:motion}a and Fig. \ref{fig:BS_method}).\medskip  

\begin{figure*}
\begin{center}
\includegraphics*[width=\linewidth, keepaspectratio=true, draft=\ddst]{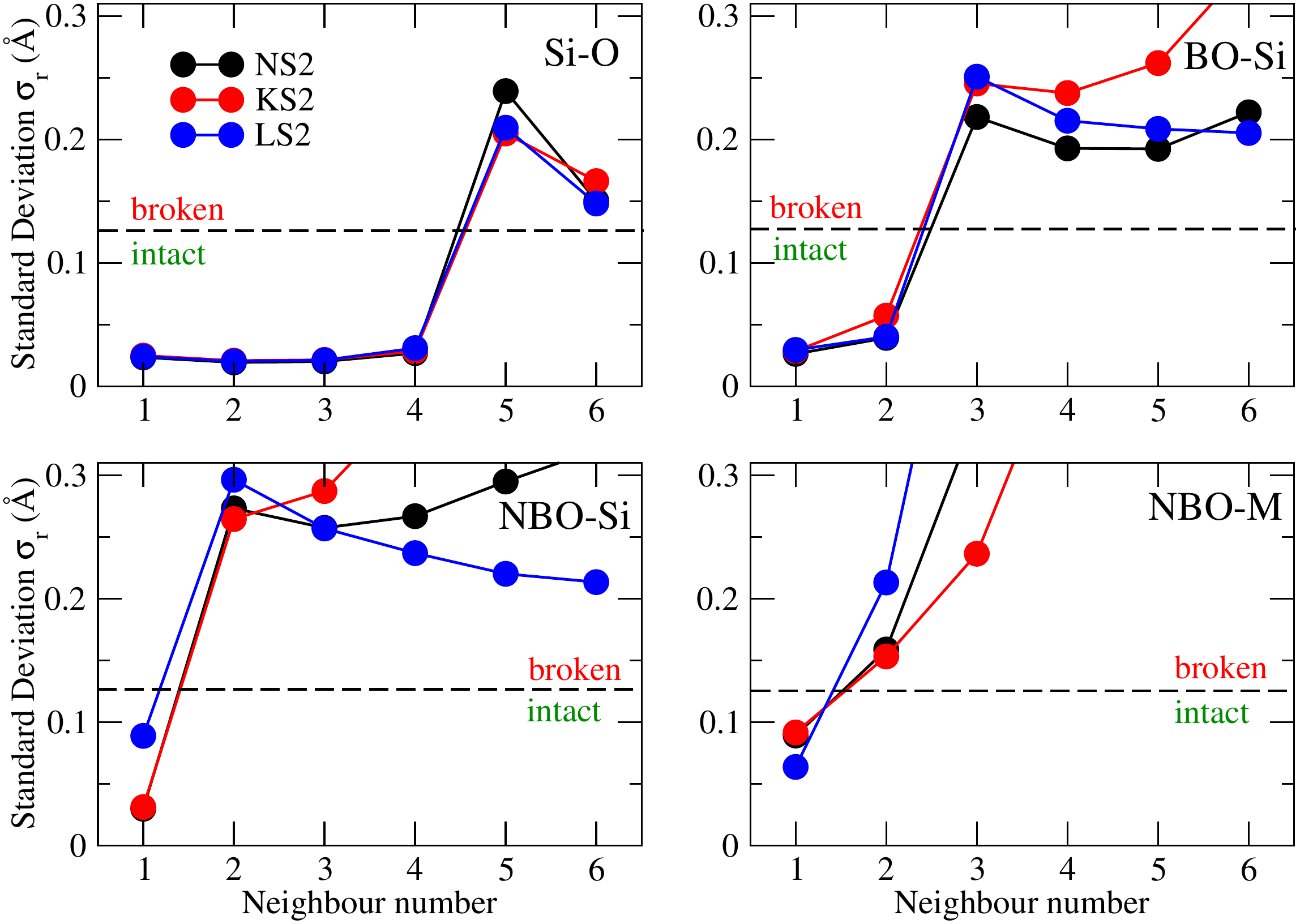}
\end{center}
\caption{Radial standard deviations $\sigma_r$ at 300K for selected pairs Si-O, BO-Si, NBO-Si and NBO-M centered neighbor distributions as a function of the neighbor number for the three alkali silicates NS2, KS2 and LS2. The broken horizontal line is an approximate limit between intact and broken constraints (see text for details).}
\label{fig:BS-300K}
\end{figure*}

Fig. \ref{fig:BS-300K} shows the standard deviations $\sigma_r$ for each relevant pair of atoms in the three alkali silicates. Note that O atoms have been splitted into bridging oxygens (BOs, connecting 2 Si tetrahedra) and non-bridging oxygens (NBOs, in the vicinity of an alkali cation) since these two species  have a different environment (and thus do not experience the same constraints). Results show a clear gap between intact (low  $\sigma_r$) and broken (large  $\sigma_r$) constraints for each glass at least for most of the species involved. As expected, Si atoms are associated to 4 BS constraints with the 4 oxygens of the tetrahedron. BOs and NBOs respectively show 2 and 1 active BS constraints with Si neighbors. The BS constraint between the NBO and the alkali atom M has a somewhat higher standard deviation ($\sigma_r \sim 0.1 \mathring{A}$) but it is still low compared to the other contributions arising from the next shell of neighbors ($\sigma_r \sim 0.2 \mathring{A}$). This suggests that the limit between intact and broken constraints should be taken between 0.1 and 0.15 $\mathring{A}$. Note that, if the 3 alkali silicates show the same general resort, associated constraints are not equivalent. In particular, the NBO-M and NBO-Si BS constraints are found to be respectively stronger and weaker in lithium silicate than in the 2 other silicates.

\subsection{Bond-Bending (BB) constraints}

\begin{figure*}
\begin{center}
\includegraphics*[width=0.6\linewidth, keepaspectratio=true, draft=\ddst]{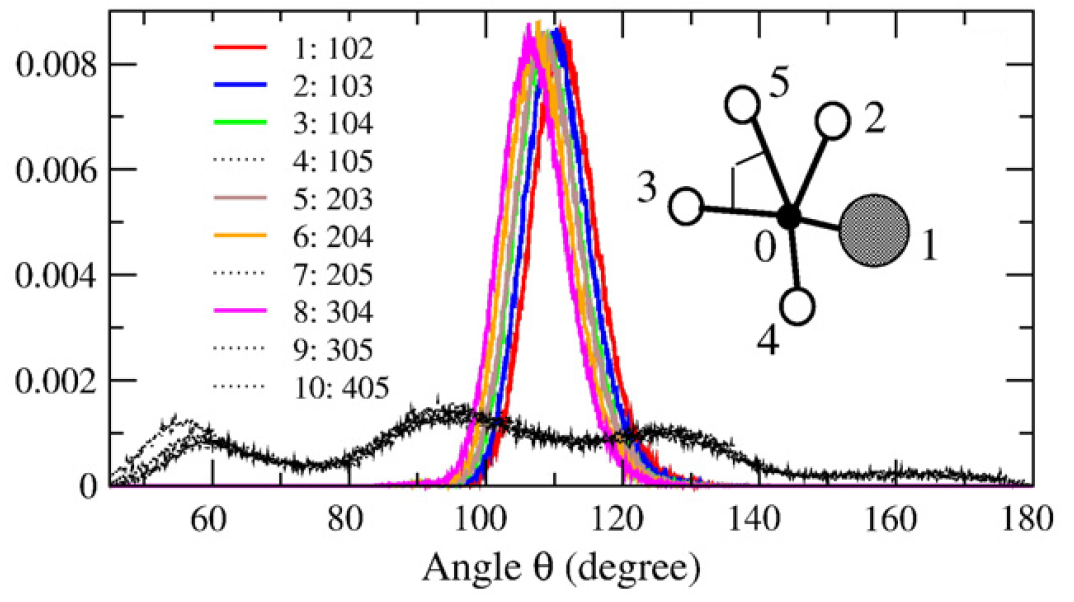}
\end{center}
\caption{Ten Si-centered partial bond angle distribution of a NS2 silicate glass for an arbitrary N. The colored curves correspond to distributions having a standard deviation $\sigma_{\theta}$ lower than 17$\textdegree$ (see also Fig. \ref{fig:BB-300K}). Other distributions are represented by broken lines. A generic view of a molecule used for the PBAD algorithm is shown : from the selection of a central atom 0, and for a given number of neighbors (here N=5), one computed all possible bond angle distributions between sets of neighbors (e.g. the marked 305).}
\label{fig:BB_method}
\end{figure*}

We have evaluated the number of BB constraints in the same fashion, by defining the partial bond angle distribution (PBADs), i.e. the distributions of the 15 angles (102, 103,...106, 203,..506) formed by a central atom 0 and each neighbors (1, 2,...6). The standard deviations $\sigma_{\theta}$ of each PBAD give a quantitative estimate of the angular excursion around the mean value, thus providing a measure of the strength of the associated bond-bending constraints. Once again, broad distributions (large $\sigma_{\theta}$) are associated to broken BB constraints whereas sharp distribution (low $\sigma_{\theta}$) to active BB constraints (see Fig. \ref{fig:motion}b and Fig. \ref{fig:BB_method}).\medskip  

\begin{figure*}
\begin{center}
\includegraphics*[width=\linewidth, keepaspectratio=true, draft=\ddst]{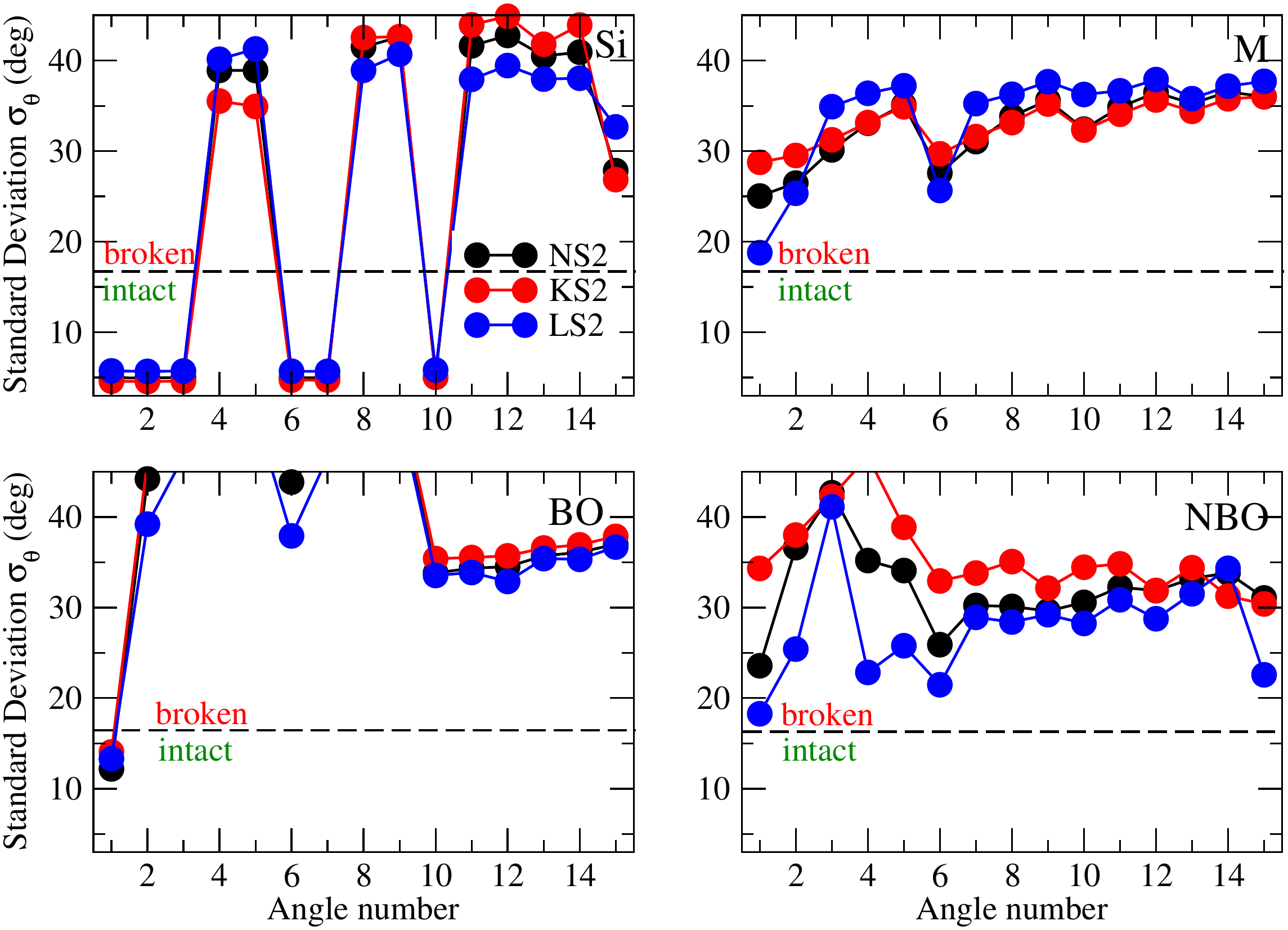}
\end{center}
\caption{Standard deviations $\sigma_{theta}$ of Si-, BO-, NBO- and M-centered (M=Na,K,Li) partial bond angle distributions as a function of the angle number for the three alkali silicates NS2, KS2 ans LS2. The broken horizontal line represents an approximate limit between intact and broken angular constraints (see text for details).}
\label{fig:BB-300K}
\end{figure*}

As for BS constraints, Fig. \ref{fig:BB-300K} shows that there is a clear gap between intact (low  $\sigma_{\theta}$) and broken (large  $\sigma_{\theta}$) constraints. As expected, only the 6 angles of the Si tetrahedron (associated to 5 independent BB constraints) and the inter-tetrahedral Si-BO-Si angle are associated to active constraints. NBO centered angles have a larger standard deviation because of the increased angular motion that manifests from the non-directional ionic NBO-M bond and the associated BB constraint is therefore broken. Note that the definition of the quite arbitrary limit between intact and broken constraints ($\sigma_{\theta}\simeq$17$\textdegree$) can be well defined when the temperature is changed, as discussed below. The present results on BS and BB constraints are thus found to match exactly a Maxwell counting assuming the 8-N rule.

\subsection{Composition effect}

\begin{figure*}
\begin{center}
\includegraphics*[width=0.6\linewidth, keepaspectratio=true, draft=\ddst]{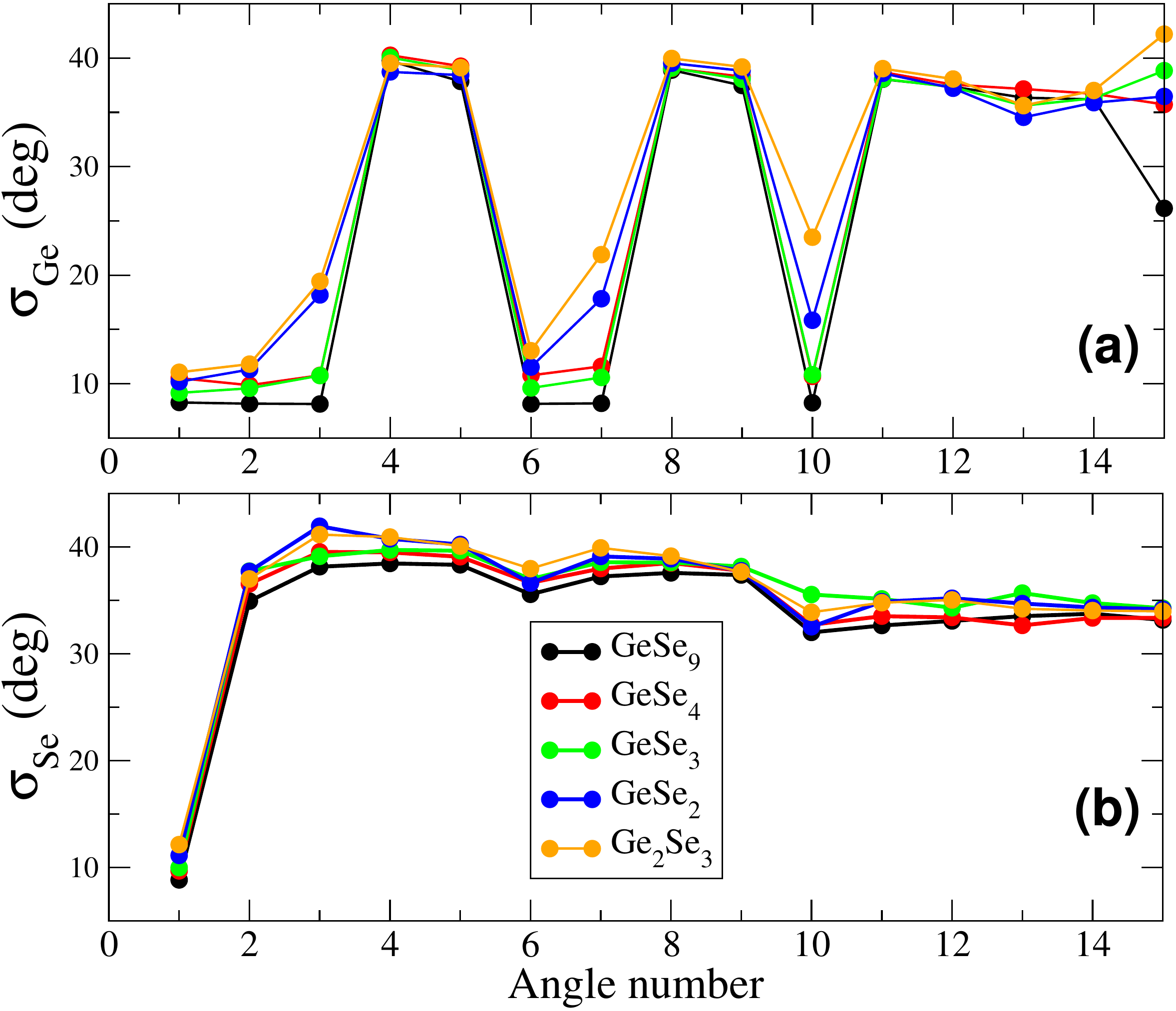}
\end{center}
\caption{Standard deviation $\sigma_{Ge}$ (a) and $\sigma_{Se}$ (b) extracted from the partial bond angle distributions (PBAD) for five selected compositions in glassy Ge$_x$Se$_{1-x}$}
\label{fig:BB-GeSe}
\end{figure*}

\begin{figure*}
\begin{center}
\includegraphics*[width=0.4\linewidth, keepaspectratio=true, draft=\ddst]{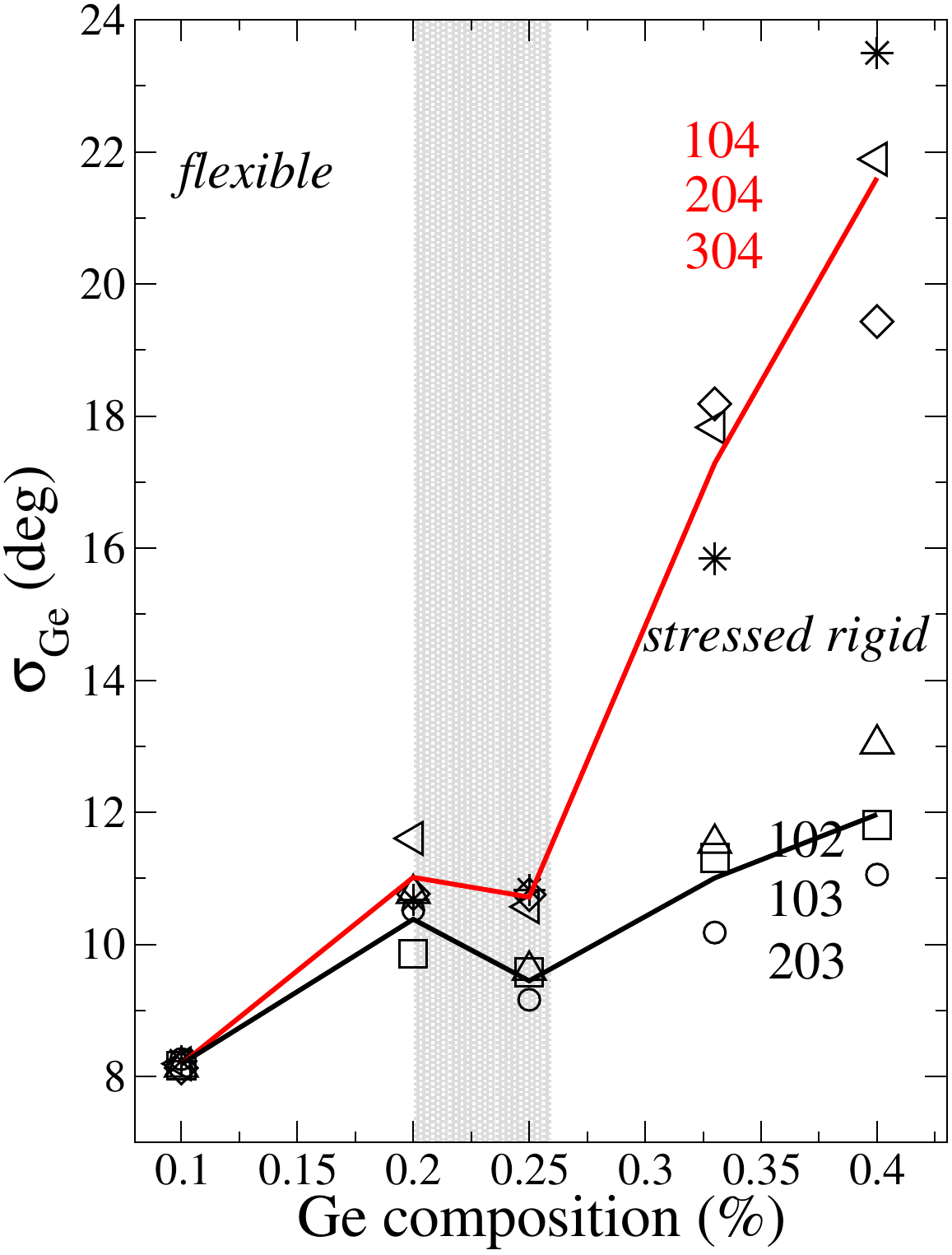}
\end{center}
\caption{Standard deviations $\sigma_{Ge}$ as a function of Ge composition, splitted into a contribution involving the fourth neighbor (red line, average of angles 104, 204 and 304) and the other contributions (black line, average of 102, 103 and 203). The shaded area corresponds to the Boolchand intermediate phase. \cite{boolchand_rigidity_2001}}
\label{fig:IP}
\end{figure*}

The same methodology has been applied on five selected compositions of Ge$_x$Se$_{1-x}$, a family of systems being flexible, rigid or isostatic depending on the composition \cite{boolchand_rigidity_2001}. Fig. \ref{fig:BB-GeSe} shows the standard deviations of the PBADs for Ge and Se atoms. As predicted by a direct Maxwell counting, the 6 angles of the Ge tetrahedron and the one of Se are found to be associated to active BB constraints. A more detailed inspection shows that there is a clear difference between flexible and intermediate compositions (x=0.10, 0.20, 0.25), having the six standard deviation $\sigma_{Ge}$ nearly equal, and stressed-rigid compositions (x=0.33, 0.40), which have different $\sigma_{Ge}$ according to the value of the neighbor, indicative that the angular motion around Ge is not symetric in stressed-rigid glasses. The deviations $\sigma_{Ge}$ involving the 4th neighbor of Ge atoms are found to be higher in the stressed-rigid phase (see Fig.\ref{fig:IP}). This highlights the fact that, in stressed-rigid systems, not all constraints can be fulfilled at the same time. These results provide the first structural evidence (from angular distributions) for the onset of stressed rigidity.

\section{Temperature-dependent constraints}

Enumerating of BS and BB constraints works well in fully connected networks at T = 0K. In practice, this situation if fulfilled as long as T is low compared to the glass transition temperature T$_g$, i.e. when thermal activation is too low to break a constraint. Extension of topological constraint theory to the liquid phase creates new scientific challenges but can lead to a better understanding of the liquid phase, based on the knowledge of the underlying low temprature network. Recently, rigidity theory has been extended by Gupta and Mauro \cite{gupta_composition_2009} to account for the temperature effect using an energy landscape approach. In this development, the behavior of each constraint is characterized by a step function q(T) so that $n_c(T)=q(T)n_c(T=0)$. At low temperature, all constraints are considered as rigid (q$\rightarrow$1). On the other hand, at high temperature, all constraints are effectively inactive (q$\rightarrow$0) because nearly all bonds can be easily broken by thermal activation. This approach has met great success by describing accurately the fragility of liquids and the glass transition temperature in different binary liquids. \cite{mauro_composition_2009, smedskjaer_prediction_2010}

\begin{figure*}
\begin{center}
\includegraphics*[width=0.6\linewidth, keepaspectratio=true, draft=\ddst]{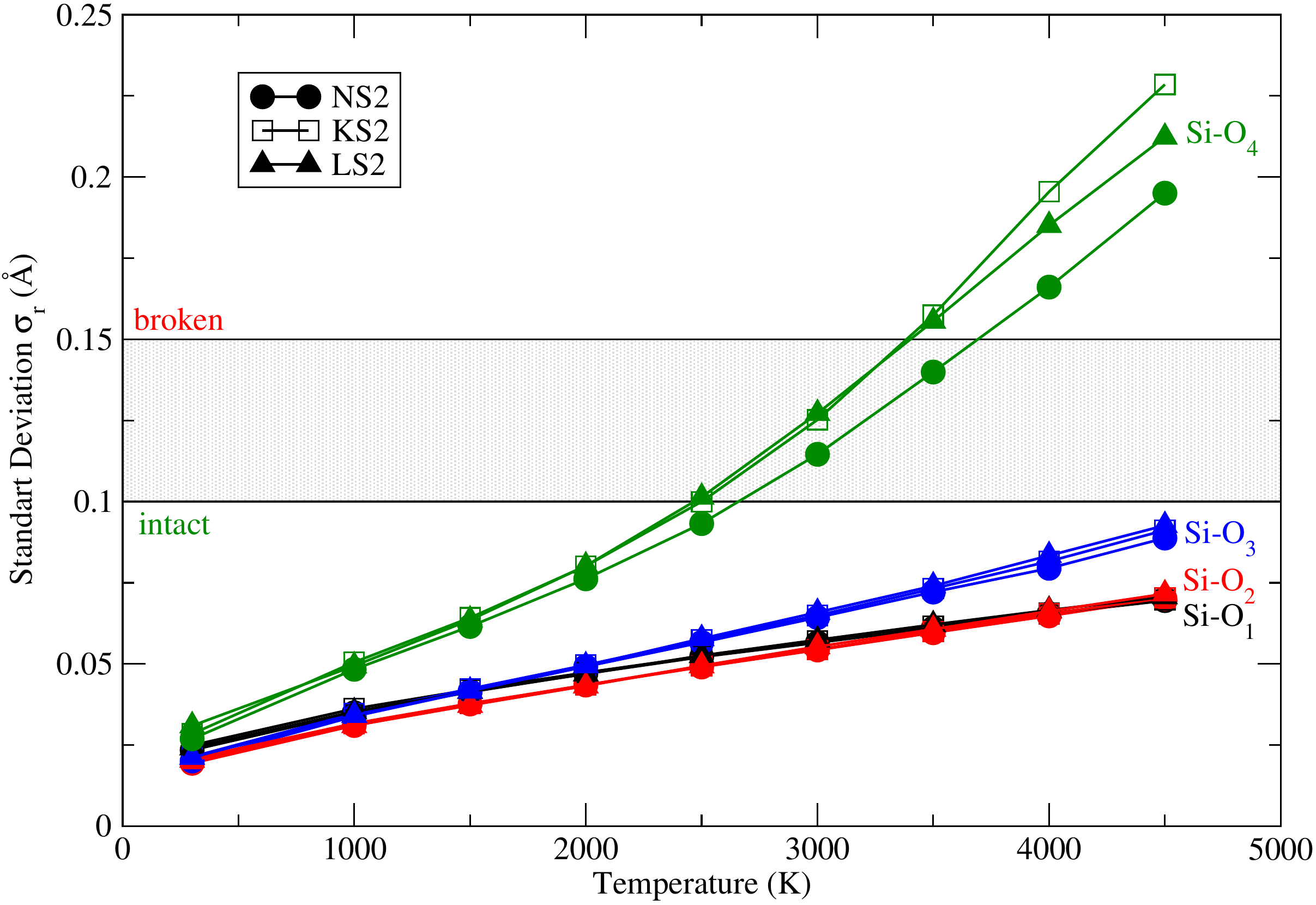}
\end{center}
\caption{Radial standard deviation $\sigma_r$ for the first neighbors of Si in NS2, KS2 and LS2. The central area indicates the region where constraints should become broken, on the basis of Fig. \ref{fig:BS-300K}.}
\label{fig:BS-SO}
\end{figure*}

We have used Molecular Dynamics to derive a physical basis for the function q(T) and, more generally, to provide an atomic scale picture to account for the behavior of the number of rigid constraints with temperature. \cite{bauchy_atomic_2011} Using the same methodology than in glasses permits to track the behavior of each constraint with temperature by following the standard deviations $\sigma_r(T)$ and $\sigma_{\theta}(T)$. Fig. \ref{fig:BS-SO} shows the radial excursion of the four O atoms around central Si for the three previously presented alkali silicates. The limit between intact and broken constraints can be approximated relying on the limit that was found in the glass (see section 2.2). These results show that the BS constraint involving the fourth oxygen neighbor becomes broken around 3000K, while the constraints coming from the three other oxygen atoms remain intact until T=4500K.\medskip 

\begin{figure*}
\begin{center}
\includegraphics*[width=0.5\linewidth, keepaspectratio=true, draft=\ddst]{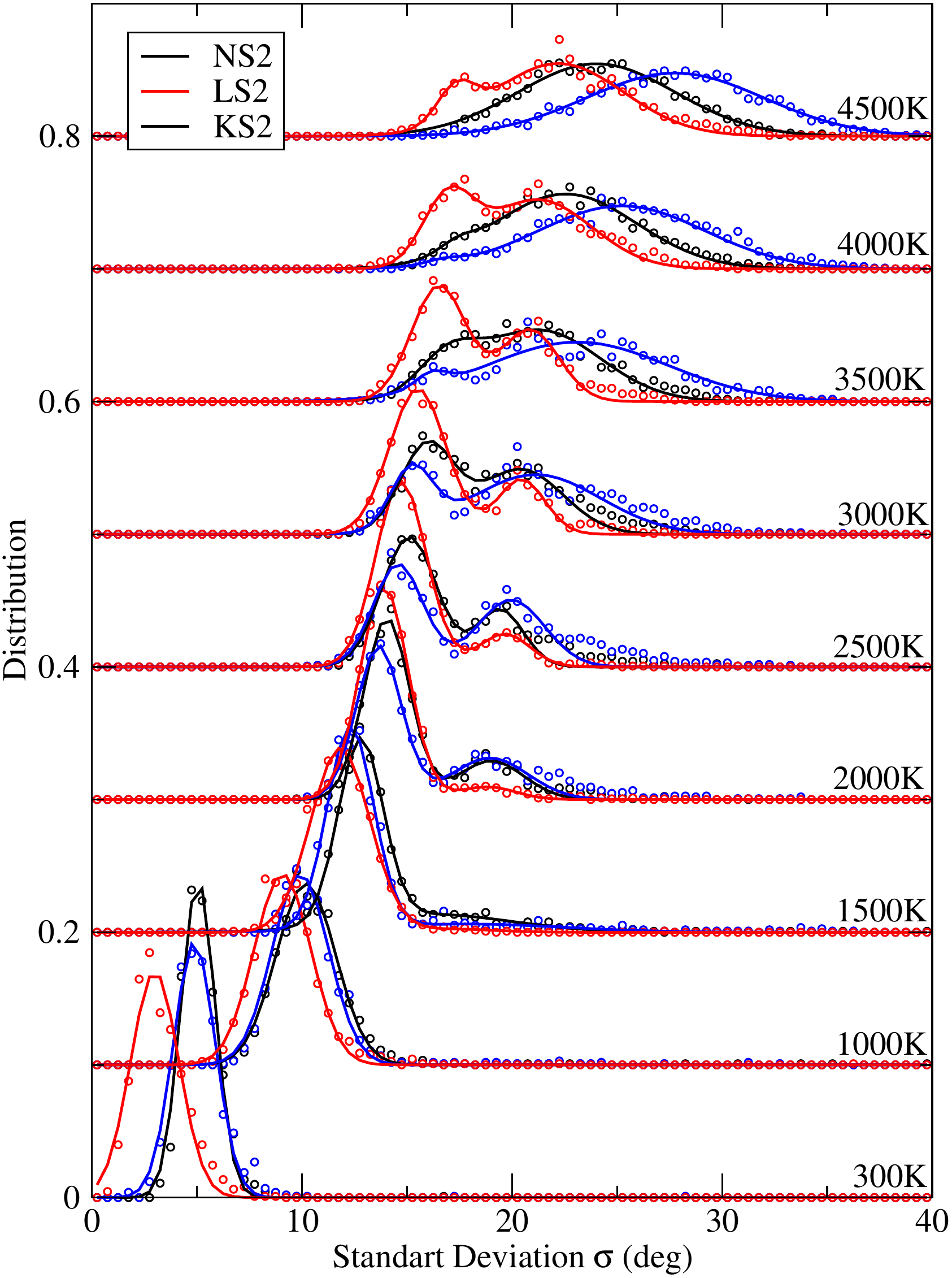}
\end{center}
\caption{Behavior of the BO central centered standard deviation $\sigma_{\theta}$ distributions for various temperature in NS2, KS2 and LS2 liquids. Bimodal distributions in the glass transition region indicate the coexistence of intact (low $\sigma_{\theta}$) and broken (large $\sigma_{\theta}$) BB constraints. Double Gaussian fits are shown on the figure.}
\label{fig:bimodale}
\end{figure*}

\begin{figure*}
\begin{center}
\includegraphics*[width=0.6\linewidth, keepaspectratio=true, draft=\ddst]{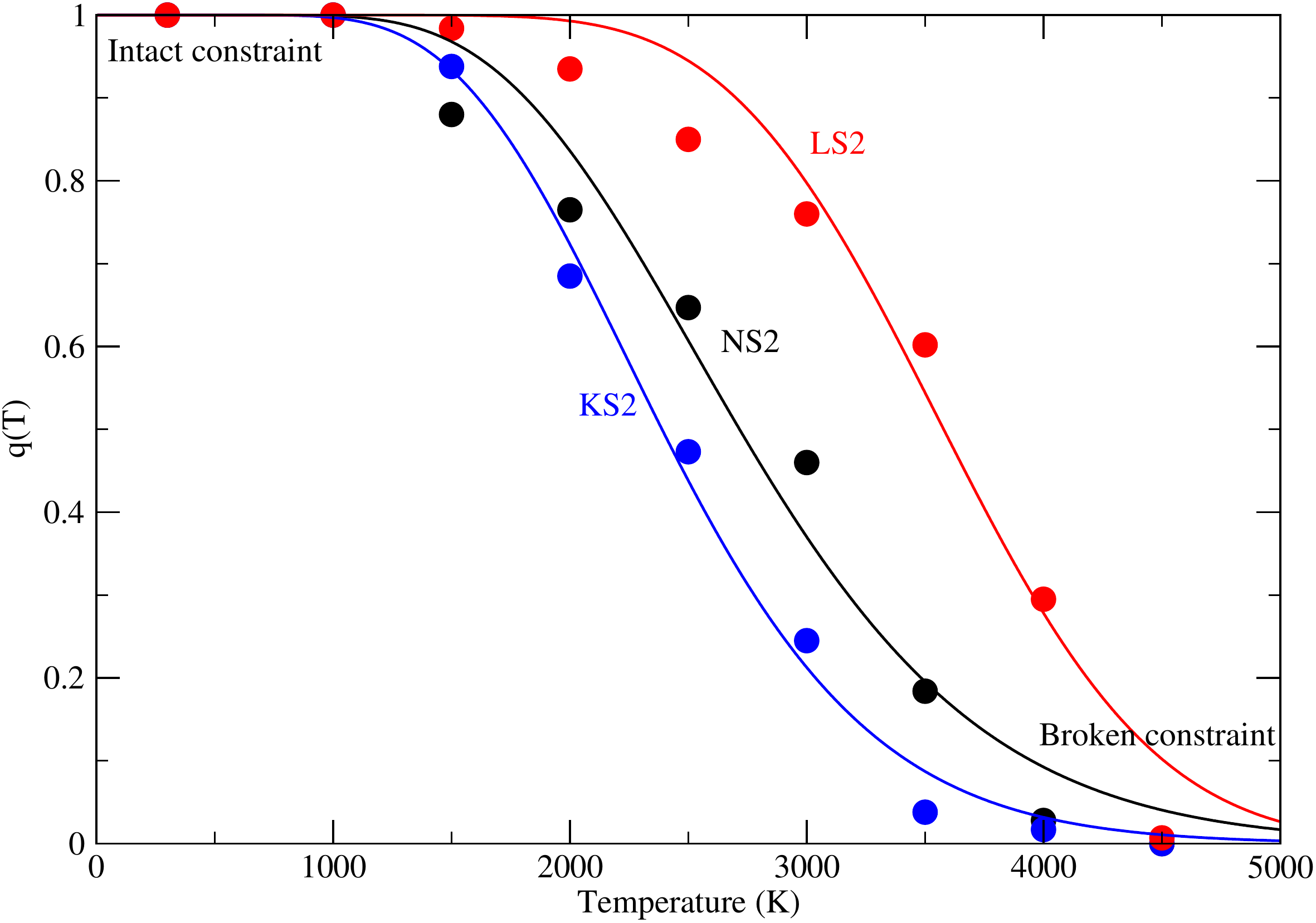}
\end{center}
\caption{Fraction q(T) of intact BO BB constraint as a function of temperature. The solid curves are fits using a formula proposed by Mauro and Gupta.}
\label{fig:q}
\end{figure*}

One can use the same method to track the behavior of BB constraints. However, to gain deeper insights into which constraints are relevant (low $\sigma$, intact) from those who are irrelevant (large $\sigma$, broken), we have analyzed the angular excursion of each individual BO by computing the angular standard deviation $\sigma_{\theta}$ of each BO in the system along the simulation time. A distribution of all the computed individual $\sigma_{\theta}$ is therefore obtained at each temperature. Fig. \ref{fig:bimodale} shows such distributions for temperature in the 300K-4500K range. At 300K, all BB constraints are intact (rather sharp distribution centered at low $\sigma_{\theta}$ value). On the other hand, at 4500K, all BB constraints must be broken because of thermal activation (as indicated by the broad distribution centered at large $\sigma_{\theta}$ value). Around T$_g$, the liquid displays a bimodal distribution, with one contribution corresponding to broken constraints (large $\sigma_{\theta}$) and the other, arising from the low temperature system, to intact constraints. The latter progressively disappears as the temperature increases. Note that this feature is also observed for the Si angle distributions. Using a double Gaussian fit, the fraction q(T) of intact BB constraints has been computed and is displayed on Fig. \ref{fig:q} for the three investigated liquids. These results exhibit the clear dichotomy between intact and broken constraints, thus offering an atomic scale foundation to count relevant constraints and to use rigidity theory on a firm basis. Finally, we have found that the approximate limit between broken and intact constraints ($\sigma_{\theta}\simeq$17$\textdegree$ on Fig. \ref{fig:BB-300K}) can be defined from the inspection of Fig. \ref{fig:bimodale}, the minimum value in the bimodal distribution allowing to define the broken-intact boundary.

\section{Pressure effect}

As composition and temperature, it is obvious that pressure can induce a rigidity transition. Indeed, it is well-known that pressure tends to increase coordination numbers in silicates and, thus, to repolymerize the network and to create new constraints \cite{trachenko_first-principles_2008}. In a preliminary work, we have investigated the transport properties (diffusion and viscosity) in a NS2 liquid at T=2000K. \cite{bauchy_pockets_2011, bauchy_viscosity_2013} \medskip  

\begin{figure*}
\begin{center}
\includegraphics*[width=0.8\linewidth, keepaspectratio=true, draft=\ddst]{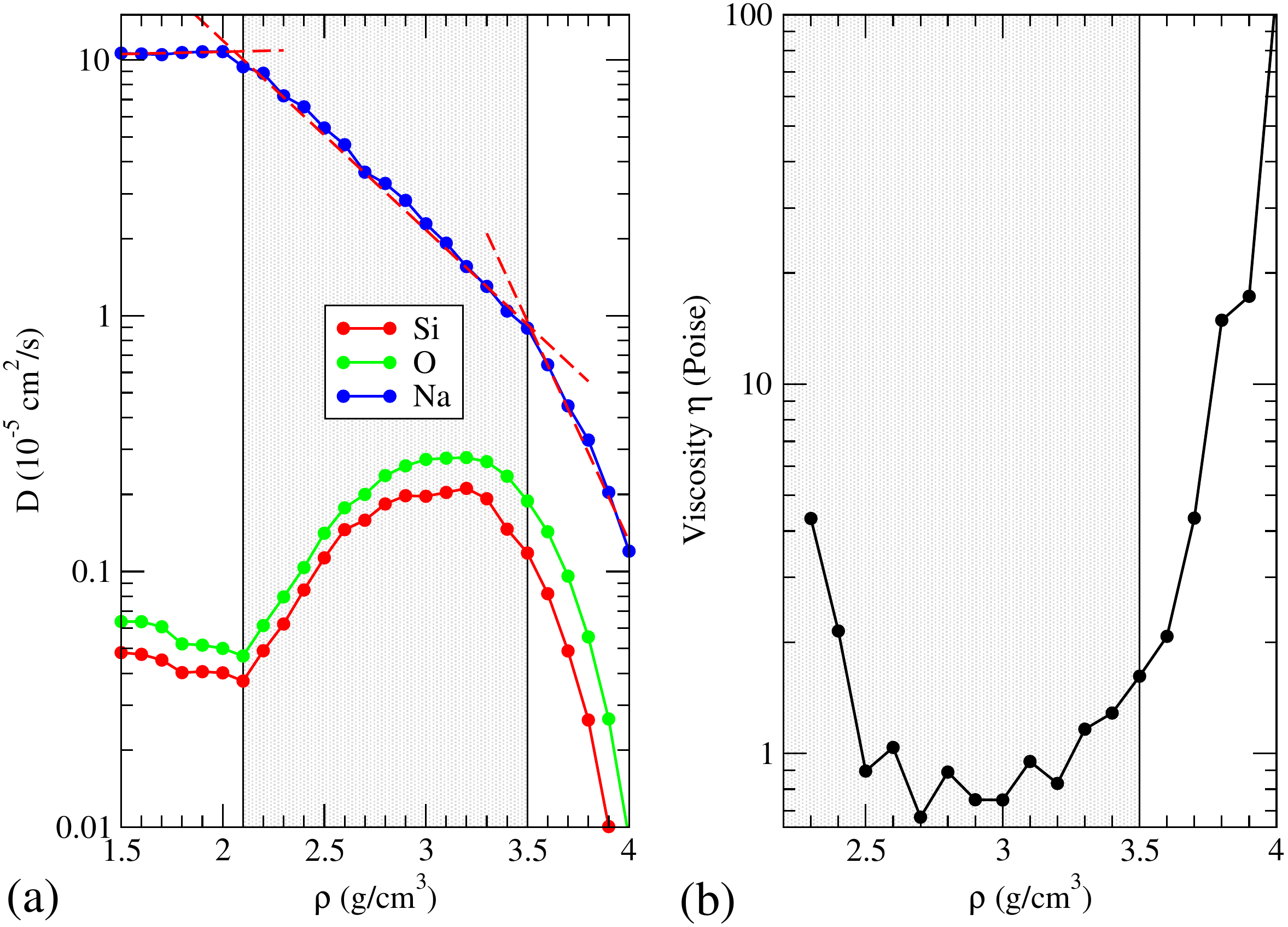}
\end{center}
\caption{(a) Silicon, Oxygen and Sodium diffusion constants of a NS2 liquid at T=2000K as a function of pressure. (b) Simulated viscosity of a NS2 liquid at T=2000K as a function of density.}
\label{fig:viscodif}
\end{figure*}

As shown on Fig. \ref{fig:viscodif}a, results show that 3 regions can be observed in the diffusion of sodium atoms. In the first regime, at low density, the diffusion constant of sodium atoms $D_{Na}$ hardly depends on the density. In the intermediate regime, between $\rho$ = 2.1 and 3.5 g/cm$^3$, $D_{Na}$ decreases with the density. Finally, in the third regime, $D_{Na}$ decreases even more rapidly. Inside the intermediate regime, we observe a maximum of the diffusion of O and Si network forming atoms, similar to the well-known example of water or silica \cite{errington_relationship_2001, shell_molecular_2002}, and the viscosity is found to display a deep minimum (see Fig. \ref{fig:viscodif}b).\medskip  

\begin{figure*}
\begin{center}
\includegraphics*[width=0.5\linewidth, keepaspectratio=true, draft=\ddst]{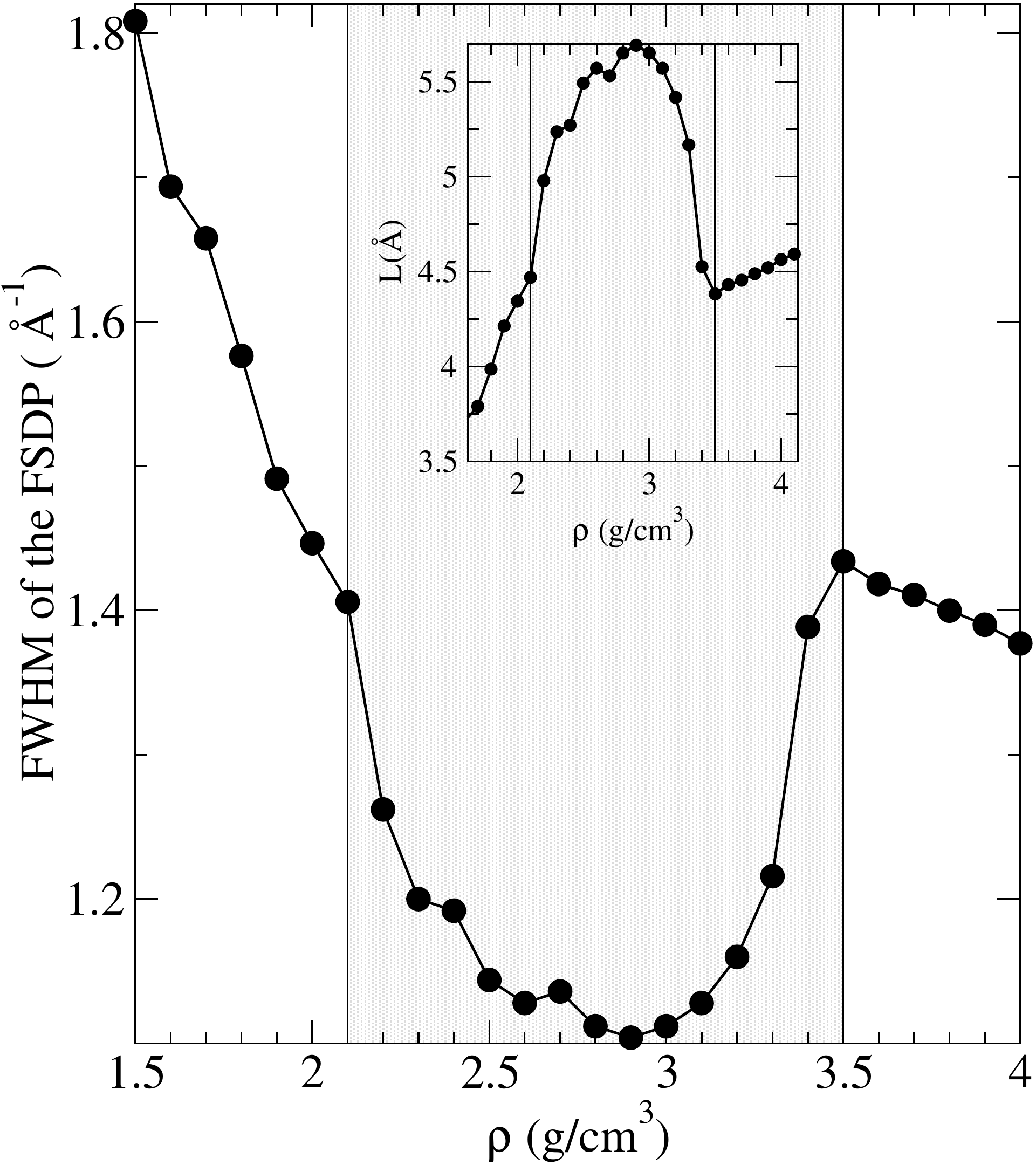}
\end{center}
\caption{Computed FWHM of the FSDP as a function of density. The shaded area corresponds to the density window from the diffusion. The insert shows the corresponding correlation length L=2 $ \pi  $/FWHM.}
\label{fig:FWHM}
\end{figure*}

Looking for a structural signature of this density window, we have computed the neutron total structure factor $S_N(Q)$. Fig. \ref{fig:FWHM} shows the full width at half maximum (FWHM) of the first diffraction peak (FSDP). Interestingly, it exhibits a density window which correlates very well with the window found from the diffusion. Note that the FWHM of the FSDP can be associated to a correlation length L$=2\pi/$FWHM which is therefore found to be maximum in this window.\medskip  

We are now in position to study combined effects with pressure and composition. These will be considered for the completion of the PhD thesis.
\cite{bauchy_structural_2012, bauchy_atomic_2011, bauchy_topological_2012, bauchy_compositional_2013, bauchy_angular_2011, bauchy_viscosity_2013, bauchy_structure_2013, bauchy_structural_2014, micoulaut_anomalies_2013, bauchy_percolative_2013, bauchy_structure_2013-1, bauchy_structural_2014-1, bauchy_contraintes_2012, micoulaut_structure_2013, bauchy_densified_2015, micoulaut_topological_2015, bauchy_pockets_2011, bauchy_angular_2010, bauchy_transport_2013}

\end{document}